\documentstyle[twoside,12pt]{article}

\font\mf=cmr8

\newcommand{\ph}{\phantom }

\newcommand{\q}{{\bf q}}
\newcommand{\r}{{\bf r}}

\newcommand{\1}{{\bf 1}}

\begin{document}

\title {\bf Replica symmetry breaking\\ in\\ attractor neural network models\\}

\author{ \\ {} \\
       Helmut Steffan\thanks{supported by Sonderforschungsbereich 123
'Stochastische Mathematische Modelle' }$\ph{a}$ and Reimer
K\"uhn\thanks{supported by a Heisenberg
 fellowship}\\ {} \\ }

\date
      {
          Institut f\"ur Theoretische Physik   \\
       Ruprecht-Karls-Universit\"at Heidelberg \\
                    Philosophenweg 19\\
             69120 Heidelberg, Germany\\
      }
  \vspace {3ex}

\maketitle

\begin{center}
Submitted to Zeitschrift f\"ur Physik B
\end{center}
  \vspace {3ex}

\begin{abstract}
\noindent {The phenomenon of replica symmetry breaking is investigated for the
retrieval phases of Hopfield-type network models. The basic calculation is done
for the generalized  version of the standard model introduced by Horner
[1] and   by Perez-Vicente and Amit [2] which can exhibit low mean
levels of  neural activity. For a mean activity $\bar a =1/2$ the Hopfield
model is recovered. In this case, surprisingly enough, we cannot confirm the
well known one step replica symmetry breaking (1RSB) result for the storage
capacity which was presented by Crisanti, Amit and Gutfreund [3]
($\alpha_c^{\hbox{\mf 1RSB}}\simeq 0.144$). Rather, we find that  1RSB- and
2RSB-Ans\"atze yield only slightly increased  capacities as compared to the
replica symmetric value ($\alpha_c^{\hbox{\mf 1RSB}}\simeq 0.138\,186$  and
$\alpha_c^{\hbox{\mf 2RSB}}\simeq 0.138\,187 $ compared to $\alpha_c^{\hbox{\mf
RS}}\simeq 0.137\,905$), significantly smaller also than the value
$\alpha_c^{\hbox{\mf sim}} = 0.145\pm 0.009$ reported from simulation studies.
These values still lie within the recently discovered reentrant phase [4]. We
conjecture that in the infinite Parisi-scheme the reentrant behaviour
disappears as is the case in the SK-spin-glass model
(Parisi--Toulouse-hypothesis). The
same qualitative results are obtained in the low activity range.}
\end{abstract}
\vfill\eject
\section{Introduction}
The Hopfield model of a neural network [5] is at present  considered to be well
understood. By suitably adopting mean field theory of spin glass like systems
[6], Amit et al. [7] were able to compute its phase diagram in a replica
symmetric (RS) approximation. In particular, they found the $T=0$ storage
capacity of the model to be $\alpha_c^{\hbox{\mf RS}}\simeq 0.138$ in the RS
framework. From simulations, they obtained a slightly larger value
$\alpha_c^{\hbox{\mf sim}}=0.145\pm0.009$ [7], and they conjectured that the
origin of such a discrepancy might be put down to effects of replica symmetry
breaking (RSB).

A subsequent one-step replica symmetry breaking (1RSB) analysis of Crisanti et
al. [3] did indeed yield $\alpha_c^{\hbox{\mf 1RSB}}\simeq 0.144$ in very good
agreement with the earlier simulation results, so that the question of the
storage capacity of this model appeared to have been settled: effects of RSB
are found to be small and 1RSB corrections seem to bridge the gap between RS
results and those of numerical simulations. Large scale simulations by Kohring
[8] seemed to confirm the overall picture.

For generalized Hopfield-type models adapted to store an ensemble of
low-activity patterns [1,2], deviations between results of RS mean field
analysis and simulation data were found to be much stronger. For instance, at
an intermediate level $\bar a =0.1$ for the mean activity, considered to be in
a neurophysiologically acceptable range, it was found that
$\alpha_c^{\hbox{\mf RS}}\simeq 0.483$ [1], whereas simulations [9] yielded
$\alpha_c^{\hbox{\mf sim}}=0.585\pm 0.01$, i.e. a discrepancy of roughly 20 \%
as opposed to a 4 \% discrepancy in the standard model.

The question then arises, whether in such a situation, a 1RSB analysis would
likewise be sufficient to close the gap between RS results and estimates
obtained from Monte-Carlo simulations, or whether -- on contrary -- further
steps in Parisi's approximating scheme [10] or even the full hierarchical
scheme of RSB [11] were needed to explain the numerical results.

In the present paper, we have addressed this question, and we have obtained
ans\-wers in completly unanticipated directions, as follows. We have performed
1RSB and 2RSB analyses for general Hopfield-type models storing ensembles of
low-activity patterns [1],[2]. For an activity $\bar a =1/2$, these
models are equivalent to Hopfield's standard model [5], and our 1RSB results
should therefore merge with those of Crisanti et al. [3] as we take the
limit $\bar a \to 1/2$. It turns out that such `confirmation en route' of the
findings of Crisanti et al. could not be accomplished, and we are forced to
conclude that their often quoted value $\alpha_c^{\hbox{\mf 1RSB}}\simeq 0.144$
for the storage capacity of the standard model is in error --- in particular
since we have invited an independent check of our results, which was recently
performed by Huyghebaert [12] using the bifurcation-finding software package
`AUTO'.

Our main results for the $\bar a =1/2$ case are $\alpha_c^{\hbox{\mf 1RSB}}
\simeq 0.138\,186$, and $\alpha_c^{\hbox{\mf 2RSB}}\simeq 0.138\,187$. These
values are still lying well within the recently discovered reentrance phase of
the Hopfield model [4]. That is, 1RSB and 2RSB approximations are found
insufficient to bridge the gap between RS and simulation results. Moreover, we
shall argue on the basis of the Parisi-Toulouse hypothesis [13] appropriatly
adapted to the Hopfield model phase diagram that even the full hierarchical RSB
scheme of Parisi [11] will not yield storage capacities as high as those
reported from various simulation studies. Rather, as in the SK-model, the
reentrance phenomenon is expected to simply disappear in the full RSB solution.

This renders the problem of how to explain discrepancies between theory and
numerical experiment an open question again.

For values other than $\bar a =1/2$ we obtain qualitatively similar results.
Here, we have performed numerical evaluations only of the 1RSB theory, and we
find the 1RSB storage capacity to lie again within the RS reentrance phase (as
it should according to the Parisi-Toulouse hypothesis), $\alpha_c^{\hbox{\mf
1RSB}}\simeq 0.495$ at $\bar a=0.1$, well below the simulation result
$\alpha_c^{\hbox{\mf sim}}=0.585\pm 0.01$ [9].

Our paper is organized as follows. In Sec. 2, we introduce the generalized
Hopfield model for the storage of low-activity patterns. In Sec. 3, we compute
the replica free energy and evaluate it in the RS, and the 1RSB and 2RSB
approximations, relegating details of the derivations to appendices. In Sec. 4
we present and discuss outcomes of a numerical analysis of the 1RSB and the
2RSB approximations for the Hopfield model, and of the 1RSB approximation for
the generalized model at $\bar a =0.1$. In Sec. 5, we discuss analogies between
the Hopfield model and the SK-model with ferromagnetic anisotropy [6] to put
our results into a wider perspective. From an appropriate adaption of the
Parisi-Toulouse hypothesis to the phase diagrams of the generalized
Hopfield-type models, we conjecture that even the full hierarchical RSB
solution of these models, presumably providing their exact solution, would
yield results which are at variance with currently available simulation data. A
concluding section is devoted to discuss the state of affairs that has thus
emerged.

\section{The Generalized Hopfield Model}

In order to set the scene and to fix our notation, we shall here introduce our
variant of the generalized Hopfield model for the storage of patterns of
arbitrary activity.

Let us denote by $\bar a$ the average fraction of active nodes in each pattern
to  be stored. We take a storage prescription of a generalized Hebbian form
\begin{eqnarray}
J_{ij} = {1\over N}\sum_{\mu=1}^p   \,\xi_i^\mu\,\xi_j^\mu,\quad i\ne j,
\label{Hebb}
\end{eqnarray}
where $1\le i,j\le N$ label the neurons of the net and $1\le \mu\le p$
enumerates the patterns. For the representation of the active ($A$) and
inactive ($I$) neural states $s_i$, we take
\begin{eqnarray}
A  =   \sqrt{1-\bar a\over \bar a} ,\quad
I  = -\sqrt{\bar a\over 1-\bar a},
\label{rep}
\end{eqnarray}
given that the pattern statistics is
 \begin{eqnarray}
\xi_i^\mu=\left\{\begin{array}{l}A,\quad \hbox{with prob.  } \ph{1-}\bar a
 \\I,\quad \hbox{ with prob.  } 1-\bar a \end{array}\right.
\label{stat}
\end{eqnarray}
The representation defined by (\ref{rep}) is known to be well adapted to the
storage of patterns with statistics given in (\ref{stat}) [1],[2],
using a generalized Hebbian storage prescription as in (\ref{Hebb}), and
supplementing it with a suitable threshold $\vartheta$. Moreover, it is known
to saturate the well known Gardner bound [14]
 \begin{eqnarray}
\alpha_c\sim {1\over 2\bar a |\ln\bar a|},
\end{eqnarray}
as $\bar a\to 0$.
The asynchronous dynamics of the model, defined by
\begin{eqnarray}
s_i(t+\Delta t)=
A\,\theta\left(h_i(t)-\vartheta\right)+I\left(1-\theta\left(h_i(t)-\vartheta
\right)\right),
\label{dyn}
\end{eqnarray}
with $h_i(t)= \sum_{j} J_{ij}\,s_j(t)$ and $\theta(x)$ the usual Heaviside
function, is governed by the energy function
\begin{eqnarray}
H[s]=-{1\over 2}\sum_{i\ne j} J_{ij}\,s_i\,s_j+\vartheta\sum_{i} \,s_i.
\label{H}
\end{eqnarray}
Asynchronous Glauber dynamics, if adapted to (\ref{dyn}), converges to a Gibbs
distribution over the space of neural states that is generated by (\ref{H}).
The next section is devoted to deal with the quenched randomness in the
couplings due to the stored patterns.

\section{The Replica Free Energy}

\subsection{General Theory}

As usual, to investigate the thermodynamics of systems with quenched
randomness, one has to compute the quenched free energy $-\beta f(\beta)=
\langle\ln Z\rangle_\xi$, where $Z$ is the partition function at fixed disorder
and where $\langle\ldots\rangle_\xi$ denotes an average over disorder according
to its distribution. The computation utilizes the replica identity
\begin{eqnarray}
\langle\ln Z\rangle_\xi=\lim_{n\to 0}{1\over n}\ln \langle Z^n\rangle_\xi.
\label{repl}
\end{eqnarray}
In the details of the calculation we follow Amit et al. [7]. Assuming that
the system state has macroscopic correlations only with a finite number of
patterns, say $\xi^\nu,\quad \nu=1,\ldots,l,\quad$ we obtain the replica free
energy through standard arguments [7], by averaging over the remaining
patterns
$\xi^\mu$:
 \begin{eqnarray}
n\,f=
{1\over2}\sum_{\nu,a} ( m^\nu_a)^2 + {\alpha\over 2\beta}\ln\det(\1-\beta\q)+
\beta\alpha\sum_{a\leq b} r_{ab}\,q_{ab}-{1\over\beta}
\langle \ln\hat Z \rangle_{\xi^\nu}
\label{nf}
 \end{eqnarray}
Here $\hat Z$ is a replica partition function
  \begin{eqnarray}
   \ph{a}\!\!\!\!\!\! \hat Z=
 \sum_{\{s^a\}} \exp\Biggl\{ \beta
 \Biggl(\sum_{a}s^{a}
\left[\sum_{\nu}  m^\nu_a\, \xi^\nu   -\vartheta\right]
+ \beta\alpha\sum_{a\leq b} r_{ab}  s^{a} s^{b} -{\alpha\over 2}\sum_{a}
(s^a)^2
\Biggr)\Biggr\},
\label{zrep}
  \end{eqnarray}
corresponding to a replicated single-site Hamiltonian of the form
 \begin{eqnarray}
   \hat H=
  -\sum_{a}s^{a}
\left[\sum_{\nu}  m^\nu_a\, \xi^\nu   -\vartheta\right]
- \beta\alpha\sum_{a\leq b} r_{ab}  s^{a} s^{b} +{\alpha\over 2}\sum_{a}
(s^a)^2.
\label{hrep}
  \end{eqnarray}
We have  introduced overlaps with the condensed patterns
 \begin{eqnarray}
m^\nu_a={1\over N}\sum_{i}   \xi_i^\nu\, s_i^{a}
  \end{eqnarray}
and the matrix $\q$ of Edwards-Anderson order parameters, with elements
 \begin{eqnarray}
q_{ab} ={1\over N}\sum_{i}     s_i^{a} s_i^{b},
  \end{eqnarray}
with $1\leq a,b\leq n$ labeling the replicas. In mean--field theory, the order
parameters must satisfy the fixed point equations
\begin{eqnarray}
m^\nu_a & = &
\langle \xi^\nu \langle s^{a}\rangle\rangle_{\xi^\nu},
\quad\nu=1,\ldots,l,\quad
a=1,\ldots,n\nonumber\\
 q_{ab}  & = & \langle  \langle s^{a} s^{b}\rangle\rangle_{\xi^\nu}, \quad
1\leq a,b\leq n,
\label{Sattel}
 \end{eqnarray}
in which $\langle\dots\rangle$ without subscript denotes a Gibbs average
performed with the Hamiltonian $ \hat H$ in (\ref{hrep}) while
$\langle\dots\rangle_{\xi^\nu}$ designates an average over the condensed
patterns $\xi^\nu$. The matrix $\r$ with elements $r_{ab}$ is simply related to
the $\q$-matrix. One has
\begin{eqnarray}
\beta\, r_{aa}  = {1\over 2}\,(\1-\beta\q)^{-1}_{aa},\quad \beta\, r_{ab} =
 (\1-\beta\q)^{-1}_{ab}, \quad a<b,
\label{rSattel}
 \end{eqnarray}
with $\1$ the $n\times n$ unit matrix. It is understood that an analytic
continuation to non-integer $n$ and the $n\to 0$ limit are eventually to be
taken.

Equations (\ref{Sattel})-(\ref{rSattel}) are usually solved by making an ansatz
concerning the transformation properties of the saddle-point values of the
order parameters $m^\nu_a$ and $ q_{ab}$ under permutation of replicas.

\subsection{The replica symmetric approximation}

The first and most natural ansatz is of course that   exhibiting complete
replica symmetry (RS):
 \begin{eqnarray}
   m_a^\nu   = m^\nu,\quad  q_{aa}=\hat q,\quad q_{ab}   = q, \quad a\neq b,
\label{repsym}
  \end{eqnarray}
the replica symmetry of $\q$ being inherited by $\r$ due to (\ref{rSattel}).
This ansatz allows for an easy evaluation of all terms appearing in (\ref{nf}),
(\ref{Sattel}) and (\ref{rSattel}), as well as for an analytic continuation of
the results to $n\to 0$. As the $n\to 0$ limit is taken one gets
 \begin{eqnarray}
f & = &
{1\over 2}\sum_{\nu} ( m^\nu)^2 -{1\over\beta}
\langle \ln\tilde Z \rangle_{z,\xi^\nu} \nonumber\\
& & + {\alpha\over 2}\left({1\over\beta}\ln\left(1-\beta(\hat
q-q)\right)-{q\over 1-\beta(\hat q-q)}+2\beta\hat r\hat q-\beta r q\right),
 \label{frepsym}
 \end{eqnarray}
where $ m^\nu$, $\hat q$ and $q$ satisfy
  \begin{eqnarray}
 m^\nu & = & \langle\, \xi^\nu  \langle s\rangle\,\rangle_{z,\xi^\nu}
\nonumber\\
\hat q & = & \langle \, \langle s^2\rangle\,\rangle_{z,\xi^\nu}
\nonumber\\
q & = & \langle \,\langle s\rangle^2  \, \rangle_{z,\xi^\nu} \nonumber\\
 \end{eqnarray}
Here $\langle \ldots\rangle$ without subscript denotes a `thermal average'
performed over the Gibbs distribution generated by the single-site Hamiltonian
\begin{eqnarray}
    \tilde H  =
  - s\biggl[ \sum_{\nu}  m^\nu\, \xi^\nu  -\vartheta+\sqrt{\alpha r}\,z\biggr]
- s^2 {\alpha\over 2}\left(\beta(2\hat r-r)-1\right),
   \end{eqnarray}
while $ \tilde Z$ is the corresponding partition function, and $\langle
\ldots\rangle_{z,\xi^\nu}$ a combined average over a zero mean unit variance
Gaussian $z$ and the $\xi^\nu$ according to their distribution. Moreover, we
have
\begin{equation}
r={q\over [1-\beta(\hat q - q)]^2} \qquad  \beta(2\hat r - r)={1 \over
1-\beta(\hat q - q)}\ .
\end{equation}
The computation leading to these equations are standard, and we shall not
document them here.

The RS solution fails to be thermodynamically stable as the temperature is
lowered through the AT-line [15], given by
\begin{eqnarray}
  T^2={ \alpha \over (1- C)^2 }\, \langle( \langle s^2\rangle- \langle
s\rangle^2)^2\, \rangle_{z,\xi},
 \label{AT}
 \end{eqnarray}
where we have introduced the `response parameter' $C=\beta\,(\hat q-q)$. The RS
Hopfield model results are recovered by taking $\vartheta=0$ and $A=-I=1$ in
(\ref{frepsym})-(\ref{AT}).

To improve upon the results of the RS approximation in the region where RS is
known to be broken according to the AT criterion [15], one can follow Parisi's
scheme of hierarchical replica symmetry breaking [10,11]. In what follows, we
present  the first two steps of Parisi's approximations for the generalized
Hopfield model. Details of the calculation are to be found in appendices A and
B for the 1RSB and the 2RSB approximations respectively.

\subsection{The 1RSB approximation}

In the 1RSB ansatz, one assumes that the overlaps $ m_a^\nu$ still exhibit the
full invariance with respect to permutations of replicas,
\begin{eqnarray}
 m_a^\nu   = m^\nu,\quad a=1,\ldots,n,
 \end{eqnarray}
whereas the Edwards-Anderson matrix $\q$ acquires the following structure,
 \begin{eqnarray}
 q_{ab}  & = & \left\{ \begin{array}{l} \hat q,\quad\ph{aaaaa} a=b,
 \\ q_1, \quad |a-b|\leq m,
\\ q_0, \quad\ph{aaa}\hbox{otherwise}. \end{array}\right.
  \label{1rsbansatz}
  \end{eqnarray}
Here $m<n$ is a partitioning parameter that is to be determined from a
stationarity condition for the free energy. Formally, we may express $\q$ in
terms of a tensor product structure
\begin{eqnarray}
{\bf q}=(\hat q-q_1)\,{\bf 1}_n+(q_1-q_0)\,
{\bf 1}_{n\over m}\otimes{\bf e}_m{\bf e}_m^{\rm T}+ q_0\,{\bf e}_n{\bf
e}_n^{\rm T},
\label{q1rsb}
\end{eqnarray}
where ${\bf 1}_k$ denotes a $k$-dimensional unit matrix and ${\bf e}_k^{\rm T}
=(1,1,\ldots,1)$ a transposed column vector with $k$ elements identical
$1$, that is, ${\bf e}_k{\bf e}_k^{\rm T}$ is nothing but a $k\times k$ matrix
completely filled with one's. The matrix $\1-\beta\q$ clearly has the same type
of tensorial structure, and so does $\r$, because it is simply related with its
inverse. These observations allow a fairly straightforward
evaluation of all terms appearing in (\ref{nf})-(\ref{hrep}), as well as of a
1RSB formulation of the saddle point equations (\ref{Sattel}). We get (for
details, see appendix A)
\begin{eqnarray}
&  & f(m^\nu,q_0,q_1,\hat q,r_0,r_1,\hat r;\,m)\,=\,{1\over2} \,  \sum_{\nu}
(m^\nu)^2 -{1\over \beta m}\,\left\langle \ln  \left\langle\,\,
  \tilde Z^{m}\right\rangle_{z_1} \right\rangle_{z,\xi^\nu}
\nonumber\\
 &  &\ \  + {\alpha\over 2}\Biggl(- {q_0 \over Q_{q_0}}+
 {1\over \beta  m}\ln\left({Q_{q_0}\over Q_{q_1}}\right) +{1\over \beta }\ln
Q_{q_1} + \beta 2\hat r\hat q
 +\beta r_1 q_1 \,(m - 1)  -\beta r_0 q_0 m \Biggr)
\label{f1rsb32}
\end{eqnarray}
for the free energy, and
\begin{eqnarray}
m^\nu & = &   \left\langle \xi^\nu { \left\langle
 \tilde Z^m \langle s \rangle \right\rangle_{z_1} \over
 \left\langle \tilde Z^m\right\rangle_{z_1}}
 \right\rangle_{z,\xi^\nu} ,
\nonumber\\
q_0 & = &   \left\langle \left( { \left\langle
 \tilde Z^m \langle s \rangle \right\rangle_{z_1} \over
 \left\langle \tilde Z^m\right\rangle_{z_1} }\right)^2
 \right\rangle_{z,\xi^\nu}\ ,
\nonumber\\
q_1 & = &   \left\langle  {  \left\langle
 \tilde Z^m \langle s \rangle ^2 \right\rangle_{z_1} \over
 \left\langle \tilde Z^m\right\rangle_{z_1} }  \right\rangle_{z,\xi^\nu}\ ,
\nonumber\\
  C\equiv \beta(\hat q- q_1) & = & {1\over \sqrt{\alpha \Delta r_1}}
\left\langle\,\,{
 \left\langle
 \tilde Z^m  {d\over dz_1} \langle s \rangle  \right\rangle_{z_1} \over
 \left\langle \tilde Z^m\right\rangle_{z_1}
}\, \,\right\rangle_{z,\xi^\nu}
\label{1rsbfpe}
\end{eqnarray}
for the fixed point equations. There is an extra equation due to the stationary
condition on $f$ with respect to the partitioning parameter $m$, which can be
expressed as follows:
\begin{eqnarray}
-{1\over \beta m }\,
 \left\langle\,\,
\ln\,\left\langle \tilde Z^{m} \right\rangle_{z_1}
\,\,\right\rangle_{z,\xi^\nu}
   & = &
-{1\over \beta m }\,
  \left\langle\,\,
{\left\langle  \tilde Z^{m}\, \ln \tilde Z^{m}\,\right\rangle_{z_1}\over
\left\langle \tilde Z^{m}\right\rangle_{z_1}}\, \,\,\right\rangle_{z,\xi^\nu}
 \nonumber\\
 & & \ph{}\!\!\!\!\!\!\!\!\!\!\!\!\!\!\!\!\!\!\!\!\!  + {\alpha\over 2}\left(
  { q_0 \over Q_{q_0}}-{ q_1 \over Q_{q_1}} - {1\over \beta m} \ln
\left({Q_{q_0}\over Q_{q_1}}\right) \right)
\label{1RSBHOPF}
\end{eqnarray}
In (\ref{f1rsb32} -- \ref{1RSBHOPF}), we have introduced the auxiliary
quantities $Q_{q_i}$ defined as
\begin{eqnarray}
Q_{q_1}=1-\beta (\hat q- q_1) \quad ,\quad Q_{q_0}= Q_{q_1} - \beta m (q_1 -
q_0)\ .
\label{Qi1rsb}
\end{eqnarray}
The elements of the $\r$-matrix, are given by
\begin{eqnarray}
r_0\ph{_2} \equiv\quad\quad  \Delta r_0 & = & \,\,\,\, {\Delta q_0 \over
Q_{q_0}^2 }\ ,
\nonumber\\
r_1-r_0 \equiv \quad\quad \Delta r_1 & = &  {\Delta q_1 \over Q_{q_0}  Q_{q_1}
}
\ ,\nonumber\\
\beta (2 \hat r-r_1)  \equiv \!\!\quad \quad\beta \Delta \hat r & = & \,\,
\,\,{1 \over   Q_{q_1} }\ ,
\label{r1rsb}
 \end{eqnarray}
where $\Delta q_0=q_0$ and $\Delta q_1=q_1 - q_0$. The quantity $\tilde Z$,
finally, denotes the partition function corresponding to the single--site
Hamiltonian
\begin{eqnarray}
\tilde H = - \left[ \sum_\nu  m^\nu \xi^\nu + \sqrt{\alpha \Delta r_0} \, z +
\sqrt{\alpha \Delta r_1} \, z_1 - \vartheta \right ] s
- {\alpha\over 2} \left[ \beta \Delta \hat r -1 \right]s^2
\label{1rsbH}
\end{eqnarray}
The notation used for the averaging brackets conforms to that introduced in the
previous subsection.

The numerical solution of these equations will be discussed in Sect. 4 below.
The case of the standard Hopfield model is recovered by taking $A=-I=1$ and
$\vartheta=0$.

\subsection{The 2RSB approximation}

The 2RSB approximation is obtained from the 1RSB scheme by endowing the
$m\times m$ diagonal submatrices of $\q$ with a structure akin to that arrived
at when breaking RS for the first time in the full matrix. Formally,
\begin{eqnarray}
\q=(\hat q-q_2)\,{\bf 1}_n+
(q_2-q_1)\,{\bf 1}_{n\over m_2}\otimes{\bf e}_{m_2}{\bf e}_{m_2}^{\rm T}+
(q_1-q_0)\,{\bf 1}_{n\over m_1}\otimes{\bf e}_{m_1}{\bf e}_{m_1}^{\rm T}+
 q_0\,{\bf e}_n{\bf e}_n^{\rm T}.
 \label{q2rsbann}
\end{eqnarray}
The tensorial structure of $\q$ is inherited by $\1-\beta\,\q$ and by $\r$ for
the same reasons as in the 1RSB case. Moreover, as in the 1RSB approximation,
one keeps replica symmetry for the overlaps $m_a^\nu$. This leads to the free
energy
\begin{eqnarray}
 & &  f(m^\nu,q_0,q_1,q_2,\hat q,r_0,r_1,r_2,\hat r;\,m_1,m_2)\,=\,
{1\over2} \, \sum_\nu (m^\nu)^2 -{1\over \beta m_1}\,\left\langle
\ln \left\langle\,\, \left\langle \tilde Z^{m_2}\right\rangle_{z_2}^{m_1 \over
m_2} \right\rangle_{z_1}  \right\rangle_{z,\xi^\nu}
\nonumber\\
 &  &  \ph{aaaaaaaaa}  + {\alpha\over 2}\Biggl( - {q_0 \over Q_{q_0}}+
{1\over \beta m_1}\ln\left({Q_{q_0}\over Q_{q_1}}\right) +{1\over \beta
m_2}\ln\left({Q_{q_1}\over Q_{q_2}}\right) +{1\over \beta }\ln Q_{q_2}
\nonumber\\
& & \ph{aaaaaaaaa}  +\beta 2\hat r\hat q
 +\beta r_2 q_2 \,(m_2-1) +\beta r_1 q_1\, (m_1-m_2)-\beta r_0 q_0 m_1 \Biggr)
\label{f2rsb}
\end{eqnarray}
and the saddle point equations
\begin{eqnarray}
 m^\nu & = & \left\langle\,\, \xi^\nu \,\,{
 \left\langle\,\,
\left\langle \tilde Z^{m_2}\right\rangle_{z_2}^{m_1\over m_2}
{\left\langle \tilde Z^{m_2} \langle s \rangle \right\rangle_{z_2}\over
\left\langle \tilde Z^{m_2}\right\rangle_{z_2}}
\,\,\right\rangle_{z_1}
\over
 \left\langle\,\,\left\langle \tilde Z^{m_2}\right\rangle_{z_2}^{m_1\over m_2}
\,\,\right\rangle_{z_1}
}\,\, \right\rangle_{z,\xi^\nu}
\\
  q_0 & = & \left\langle\,\, \left( \,\,{
 \left\langle\,\,
\left\langle \tilde Z^{m_2}\right\rangle_{z_2}^{m_1\over m_2}
{\left\langle \tilde Z^{m_2} \langle s \rangle \right\rangle_{z_2}\over
\left\langle \tilde Z^{m_2}\right\rangle_{z_2}}
\,\,\right\rangle_{z_1}
\over
 \left\langle\,\,\left\langle \tilde Z^{m_2}\right\rangle_{z_2}^{m_1\over m_2}
\,\,\right\rangle_{z_1}
}\,\,\right)^2\,\, \right\rangle_{z,\xi^\nu}
\\
 q_1 & = & \left\langle\, \,{
 \left\langle\,\,
\left\langle \tilde Z^{m_2}\right\rangle_{z_2}^{m_1\over m_2}\, \left( \,
{\left\langle \tilde Z^{m_2} \langle s \rangle \right\rangle_{z_2}\over
\left\langle \tilde Z^{m_2}\right\rangle_{z_2}}\,\right)^2\,
\,\,\right\rangle_{z_1}
\over
 \left\langle\,\,\left\langle \tilde Z^{m_2}\right\rangle_{z_2}^{m_1\over m_2}
\,\,\right\rangle_{z_1}
}\, \, \right\rangle_{z,\xi^\nu}
\\
 q_2 & = & \left\langle\, \,{
 \left\langle\,\,
\left\langle \tilde Z^{m_2}\right\rangle_{z_2}^{m_1\over m_2}\,
{\left\langle \tilde Z^{m_2} \langle s \rangle^2\,\right\rangle_{z_2}\over
\left\langle \tilde Z^{m_2}\right\rangle_{z_2}}\,
\,\,\right\rangle_{z_1}
\over
 \left\langle\,\,\left\langle \tilde Z^{m_2}\right\rangle_{z_2}^{m_1\over m_2}
\,\,\right\rangle_{z_1}
}\, \, \right\rangle_{z,\xi^\nu}
\\
C\equiv \beta(\hat q- q_2) & = & {1\over \sqrt{\alpha\Delta r_2}}\left\langle\,
\,{ \left\langle\,\,
\left\langle \tilde Z^{m_2}\right\rangle_{z_2}^{m_1\over m_2}\,
{\left\langle\tilde Z^{m_2}{d\over dz_2} \langle s
\rangle\,\right\rangle_{z_2}\over \left\langle \tilde
Z^{m_2}\right\rangle_{z_2}}\,
\,\,\right\rangle_{z_1}
\over
 \left\langle\,\,\left\langle \tilde Z^{m_2}\right\rangle_{z_2}^{m_1\over m_2}
\,\,\right\rangle_{z_1}
}\, \, \right\rangle_{z,\xi^\nu}
\label{2rsbfpe}
\end{eqnarray}
The partitioning parameters $m_1$ and $m_2$ are determined by stationarity
conditions on $f$, leading to two further fixed point equations, namely:
\begin{eqnarray}
  -{1\over \beta m_1 }\,\left\langle \ln  \left\langle\,\,
\left\langle \tilde Z^{m_2}\right\rangle_{z_2}^{m_1\over m_2}
\right\rangle_{z_1}   \right\rangle_{z,\xi^\nu}\!\!\!\!\!\!
& = & -{1\over \beta m_2 }\,
\left\langle\, \,{
 \left\langle\,\,
\left\langle \tilde Z^{m_2}\right\rangle_{z_2}^{m_1\over m_2}\,
\ln\,\left\langle \tilde Z^{m_2} \right\rangle_{z_2} \,\,\right\rangle_{z_1}
\over
 \left\langle\,\,\left\langle \tilde Z^{m_2}\right\rangle_{z_2}^{m_1\over m_2}
\,\,\right\rangle_{z_1}
}\, \, \right\rangle_{z,\xi^\nu}
\nonumber\\
& &
\ph{}\!\!\!\!\!\!\!\!\!\!\!\!\!\!\!\!\!\!\!\!\!\!\!\!\!\!\!\!\!\!\!\!\!\!\!\!
+ {\alpha\over 2}\left(  { q_0 \over Q_{q_0}}-{ q_1 \over Q_{q_1}}
-{1\over \beta m_1}\ln\left({Q_{q_0}\over Q_{q_1}}\right)  \right)
\label{stat1}
 \\
& & \ph{\Biggl(\Biggr)}\nonumber\\
-{1\over \beta m_2 }\,
\left\langle\, \,{
 \left\langle\,\,
\left\langle \tilde Z^{m_2}\right\rangle_{z_2}^{m_1\over m_2}\,
\ln\,\left\langle \tilde Z^{m_2} \right\rangle_{z_2} \,\,\right\rangle_{z_1}
\over
 \left\langle\,\,\left\langle \tilde Z^{m_2}\right\rangle_{z_2}^{m_1\over m_2}
\,\,\right\rangle_{z_1}
}\, \, \right\rangle_{z,\xi^\nu}\!\!\!\!\!\!\!\!
  & = &
-{1\over \beta m_2 }\,
\left\langle\, \,{
 \left\langle\,\,
\left\langle \tilde Z^{m_2}\right\rangle_{z_2}^{m_1\over m_2}\,
{\left\langle \tilde Z^{m_2}\, \ln\,\tilde Z^{m_2}\,\right\rangle_{z_2}\over
\left\langle \tilde Z^{m_2}\right\rangle_{z_2}}\,
\,\,\right\rangle_{z_1}
\over
 \left\langle\,\,\left\langle \tilde Z^{m_2}\right\rangle_{z_2}^{m_1\over m_2}
\,\,\right\rangle_{z_1}
}\, \, \right\rangle_{z,\xi^\nu}
\nonumber\\
 & &
\ph{}\!\!\!\!\!\!\!\!\!\!\!\!\!\!\!\!\!\!\!\!\!\!\!\!\!\!\!\!\!\!\!\!\!\!\!\!
 + {\alpha\over 2}\left( {q_1 \over Q_{q_1}}-{ q_2 \over Q_{q_2}}
-{1\over \beta m_2}\ln\left({Q_{q_1}\over Q_{q_2}}\right)  \right)
\label{stat2}
\end{eqnarray}
The $\r$ elements, finally, are algebraically related to the elements of $\q$:
\begin{eqnarray}
r_0 \ph{_2} \equiv\quad\quad  \Delta r_0 & = & \,\,\,\, {\Delta q_0 \over
Q_{q_0}^2 }\ ,
\nonumber\\
r_1-r_0  \equiv \quad\quad \Delta r_1 & = &  {\Delta q_1 \over Q_{q_0}  Q_{q_1}
}\ ,
\nonumber\\
r_2-r_1 \, \equiv\quad\quad  \Delta r_2 & = &  {\Delta q_2 \over Q_{q_1}
Q_{q_2} } \ ,
\nonumber\\
\beta (2 \hat r-r_2)  \equiv \!\!\quad \quad\beta \Delta \hat r & = & \,\,
\,\,{1 \over   Q_{q_2} }\ .
\label{r2rsb}
\end{eqnarray}
Here $\Delta q_0 = q_0$, and $\Delta q_i= q_i - q_{i-1}$ for $i=1,2$, while
\begin{eqnarray}
Q_{q_2}=1-\beta (\hat q- q_2) \ ,\quad Q_{q_1}= Q_{q_2} - \beta m_2 (q_2 -
q_1) \ ,\quad Q_{q_0}= Q_{q_1} - \beta m_1 (q_1 - q_0)\ .
\label{Qi2rsb}
\end{eqnarray}
As in the  previous subsection, $\tilde Z$ denotes a partition function
corresponding to a single--site Hamiltonian, namely
\begin{eqnarray}
\tilde H = - \left[ \sum_\nu  m^\nu \xi^\nu + \sqrt{\alpha\Delta r_0} \, z +
\sqrt{\alpha \Delta r_1} \, z_1 + \sqrt{\alpha \Delta r_2} \, z_2 - \vartheta
\right ] s - {\alpha\over 2} \left[ \beta\Delta \hat r -1 \right]s^2\ .
\label{2rsbH}
\end{eqnarray}
Conventions regarding averaging brackets are the same as before.

{}From the structure of the 1RSB and the 2RSB equations, a formulation of the
infinite RSB scheme is fairly easily obtained. Since we have not evaluated this
limit numerically, we will not reproduce the corresponding equations here. The
interested reader will find the details in [16].

\section{Results}

We have solved the fixed point equations corresponding to the RS, the 1RSB and
the 2RSB approximations for the Hopfield model with $\bar a=1/2$, as well as
those corresponding to the RS and the 1RSB approximations for the generalized
model at $\bar a=0.1$. In both cases, simulation data are available for
comparison with theoretical results [7--9]. The RS approximations reproduce
previously known results as they should.

As usual, the numerics simplifies considerably in the $T=0$ limit, because the
innermost Gaussian averages in the saddle  point equations can be performed
analytically in this limit, giving simple expressions in terms of error
functions. Moreover, it can be shown that, as this limit is taken, the
partitioning parameters $m$ and $m_1, m_2$ of the 1RSB and the 2RSB
approximations enter the theory only through the scaled combinations
$D=\beta\,m$ and $D_1=\beta\,m_1,\,D_2=\beta\,m_2$, which remain finite as the
$\beta\to\infty$ limit is taken. In the case of the 1RSB approximations, this
has already been noticed by Crisanti et al. [3].

\subsection{The standard model at $\bar a=1/2$}

In the case of the standard model, we have $A=-I=1$ and $\vartheta=0$. The full
phase diagram in the RS approximation is well known [7]. In Fig. 1 we present,
for later reference, an enlarged portion of it, exhibiting the boundary of the
retrieval phase at low temperatures, as well as its AT-line. A noteable feature
here is the reentrant behaviour signified by a backbending of the transition
line for $T\leq T(\alpha_{\hbox{\mf max}})=0.024$, previously discovered by
Naef and Canning [4]. Moreover, the AT-line below which the RS approximation
fails to be thermodynamically acceptable is seen to meet the RS phase boundary
just slightly above $T(\alpha_{\hbox{\mf max}})$, where reentrant behaviour
begins. We will return to discussion of these features in Sec.5 below.

\centerline{ }
\centerline{Fig. 1}
\centerline{ }

Knowing that replica symmetry must be broken below the AT-line, we have
analyzed the 1RSB approximation of the model. Fig. 2 shows the order parameters
$m^1$,
$q_1$, and $q_0$ ($\hat q=1$) at $T=0.02$, for a range of $\alpha$ values that
cross the AT-line at $\alpha_{\hbox{\mf AT}}\simeq 0.137\,6$. As is to be
expected $q_1$ and $q_0$ become different, as $\alpha$ is increased through
$\alpha_{\hbox{\mf AT}}$. The figure exhibits both stable and unstable
solutions of the order parameters. The stable and unstable branches of these
order
parameters all coalesce at $\alpha^{\hbox{\mf 1RSB}}_c(T)\simeq 0.138\,19$,
signifying the tangent-bifurcation that marks the boundary of the retrieval
phase at this temperature in the 1RSB approximation.
At zero temperature, $T=0$, we find the following critical parameters
\begin{eqnarray}
 \alpha_c & \simeq &   0.138\,186\,489\,5\nonumber\\
m^1 & \simeq & 0.966\,77\nonumber\\
q_0 & \simeq & 0.996\,48\nonumber\\
C & \simeq & 0.052\,89\nonumber\\
 D & \simeq & 36.78
\end{eqnarray}
These values deviate considerably from those previously reported by Crisanti et
al. [3]
\begin{eqnarray}
 \alpha_c   &   \simeq &  0.144\nonumber\\
   m^1 & \simeq &  0.982
\nonumber\\
 C & \simeq &  0.111
\nonumber\\
 D & \simeq &  0.03
\end{eqnarray}
which they obtained through a Monte-Carlo minimization of the free energy
function (\ref{f1rsb32}), rather than by deriving and solving the associated
fixed point approximations.

\centerline{ }
\centerline{Fig. 2}
\centerline{ }

In view of this discrepancy, we have performed several internal consistency
checks of our results. Since our expression for the free energy is the same as
in [3], a possible discrepancy could arise due to errorneous expressions for
the fixed point equations. However, our 1RSB fixed point system can in
principle also be derived from the 2RSB approximation by either taking the
limit $m_2\to
1$ or the limit $m_1\to n$. Both checks confirmed the expressions given in
(\ref{1rsbfpe}--\ref{1rsbH}). Moreover, we have some internal consistency
checks by isolating asymptotics of various integrations analytically, with
essentially no change on the results. We have also omitted the stationarity
requirement with
respect to the partitioning parameter $m$, treating $D=\beta\, m$ as a free
parameter, and computing $\alpha_c$ at $T=0$ as a  function of $D$. The result
is shown in Fig. 3. The capacity $\alpha_c(D)$ never increases beyond
$\alpha^{\hbox{\mf 1RSB}}_c(D)_{\hbox{\mf max}}\simeq 0.138 2$, approaching ---
as it should --- the replica symmetric capacity $\alpha_c^{\hbox{\mf
RS}}=0.137\,905$
as $D\to 0$ or $D\to \infty$, albeit in the latter case slowly. Even
$\alpha^{\hbox{\mf 1RSB}}_c(D)_{\hbox{\mf max}}$ is found to be slightly
smaller than $\alpha^{\hbox{\mf RS}}_c(T)_{\hbox{\mf max}}\simeq
0.138\,188\,5$. Lastly, an independent check of our results was recently
obtained by Huyghebaert [12],
using the bifurcation finding software package `AUTO', and confirming our
results to an accuracy of 9 significant digits.

\centerline{ }
\centerline{Fig. 3}
\centerline{ }

\begin{table}[h]
\begin{center}
\begin{tabular}{|l|l|l|l|l|l|l|l|l}
\hline
 &  $\ph{aaaa}\alpha_c$ &  $\ph{aaa} m^1$  &   $ \ph{a}D$ &   $ \ph{a}D_1$ &
$\ph{a} D_2$ & $\ph{aaa}f=u$ &  $\ph{aaaa}s$ \\ \hline \hline
 $\ph{1}$RS$\ph{\rm B}$ & 0.137\,905\,566 & 0.967\,417 & &  & &
-0.501\,445\,395 &  -0.001\,445 \\ \hline
 1RSB &  0.138\,186\,489 & 0.966\,777 & 36.783 &  &   & -0.501\,446\,051  &
-0.000\,104 \\ \hline
 2RSB &  0.138\,187\,733 & 0.966\,776 &   & 2.406 & 38.320  &  -0.501\,446\,125
 & -0.000\,097 \\ \hline
 \end{tabular}
 \end{center}
\hskip 1cm \parbox{14cm}{ \caption{ \label{Tab1}  \protect\small
Retrieval boundary at $T=0$ in RS, 1RSB and 2RSB approximations for the
standard model with $\bar a =1/2$. Also given are values for (free) energy $u$
and entropy $s$.}}
\end{table}

We have also considered the 2RSB approximation in  the $T\to 0$ limit. Because
of additional integrations that need to be performed numerically, these results
are inherently less precise than those  for the 1RSB scheme. They are collected
in
Tab. 1.

Note that the 2RSB approximation gives only a rather slight increase in the
$T=0$ storage capacity, which is still within the RS reentrant phase, i.e.,
$\alpha^{\hbox{\mf 2RSB}}_c < \alpha^{\hbox{\mf RS}}_c(T)_{\hbox{\mf max}}$.
Fig. 4  shows the Parisi function, as a function of the rescaled partitioning
parameters $D_i=\beta\, m_i$, in the 1RSB and the 2RSB approximations,
respectively.

\centerline{ }
\centerline{Fig. 4}
\centerline{ }

\subsection{The general model at $\bar a=0.1$}

For the general model ($\bar a$ different from $1/2$), there is no longer a
symmetry between the active and inactive neural states. In what follows we
shall exclusively deal with the $\bar a=0.1$ case, for which simulations
results are
available for comparision [9]. A non-zero threshold has to be introduced, and
it must be optimized in order to yield the largest retrieval region. In Fig. 5,
we show the RS boundary of the retrieval region for the  generalized model at
$\bar a=0.1$, for various values of the threshold $\vartheta$. The optimal
threshold,
yielding the largest $T=0$ capacity is found to be $\vartheta_{\mf opt}\simeq
1.825\,57$ for this system.

In principle, the envelope of the retrieval regions attainable with
continuously varying $\vartheta$ gives the ultimate boundary for retrieval.
This would
require $\vartheta$ to vary with temperature, $\vartheta=\vartheta(T)$. Note
the strong reentrant behaviour in case of near optimal threshold, clearly
visible in the present case {\em without} amplification.

As in the standard model, the RS solution fails to be acceptable for
$T < T_{\hbox{\mf AT}}(\alpha)$, given by (\ref{AT}), and depicted in Fig. 6
for $\vartheta=\vartheta_{\hbox{\mf opt}}$. Again, exactly as in the standard
model, the AT-line is seen to meet the RS phase boundary just slightly above
$T(\alpha_{\hbox{\mf max}})$ below which the RS phase boundary bends back to
lower values of $\alpha$. For the present case $T(\alpha_{\hbox{\mf
max}})\simeq 0.24$ at $\alpha_{\hbox{\mf max}}\simeq 0.495\,17$, whereas the
$T=0$ capacity is
\begin{eqnarray}
 \alpha_c^{\hbox{\mf RS}} & \simeq &   0.484\,15
\end{eqnarray}
We have investigated the 1RSB approximation of this model at $T=0$, with
results collected in Tab. 2.

\begin{table}[t]
\begin{center}
\begin{tabular}{|l|l|l|l|l|l|l|l|l}
\hline
 &  $\ph{aaaa}\alpha_c$ &  $\ph{aaa} m^1$  &   $ \ph{aa} C$ &      $\ph{a} D$ &
$\ph{aaa}f=u$ &  $\ph{aaaa}s$ \\ \hline \hline
 $\ph{1}$RS$\ph{\rm B}$ & 0.484\,151\,834 & 0.836\,979 & 0.176\,688   & &
-0.508\,557\,635 &  -0.004\,886 \\ \hline
 1RSB &  0.495\,030\,302 & 0.828\,196 & 0.055\,155   & 3.523  &
-0.509\,302\,481  & -0.000\,406 \\ \hline
 \end{tabular}
\end{center}
\hskip 1cm \parbox{14cm}{ \caption{ \label{Tab2}  \protect\small
Retrieval boundary at $T=0$ in the RS and the 1RSB approximations for  the
generalized model at $\bar a=0.1$ and with optimal threshold
$\vartheta=1.82557$.}}
\end{table}

\centerline{ }
\centerline{Fig. 5,6}
\centerline{ }

The $q$ values for this case are at $\hat q = q_1 \simeq 0.929\,310$ and $q_0
\simeq 0.891\,226$ compared to the replica symmetric $\hat q = q \simeq
0.924\,081$. While the relative increase in computed $T=0$ storage capacity due
to 1RSB corrections is significantly larger than in the standard model (roughly
$2\%$ as opposed to only $0.15\%$ in the standard model), it does still give a
capacity {\em within} the reentrant phase, and far below the number
$\alpha_c^{\hbox{\mf sim}}=0.585\pm 0.01$ reported from numerical simulations
[9]. Again we have checked our results by treating $D$ as an independent
parameter, not fixed by a stationarity requirement, with results qualitatively
similar to the case of the standard model; see Fig. 7.

\centerline{ }
\centerline{Fig. 7}
\centerline{ }

Though we have not analyzed the 2RSB approximation, we expect the outcome of
such an analysis to be qualitatively similar to the standard case: there will
be an additional slight increase in the $T=0$ storage capacity, but it will
still
be smaller than $\alpha_c^{\hbox{\mf RS}}(T)_{\hbox{\mf max}}$, i.e. still be
within the reentrant phase.

A final note here concerns the sharp bends in the RS phase diagrams, at
temperatures above $T(\alpha_{\hbox{\mf max}})$. They are due to the fact that
the RS retrieval phase loses stability in different directions giving way to
different frozen phases, depending on whether $\alpha < \alpha^{*}$ or
$\alpha > \alpha^{*}$, where $\alpha^{*}$ denotes the loading level at which
the sharp bend occurs. Since the non-retrieval phases are not so much of
concern to
us in the present context, we will not elaborate on this point, however. The
interested reader may consult [1] and [16] on this matter.

\section{Relation with RSB in the SK-model}

In the previous section, we have seen that 1RSB or 2RSB approximations to the
mean-field solution of generalized Hopfield-type models yield a slight increase
of the storage capacities of these models, but this increase was found to be
{\em much smaller} than probably expected from simulation results [7--9], or
previously reported [3]. In all cases , the resulting storage capacity was
found
to be smaller than that maximally attainable in the RS approximation at finite
temperature. In Sect. 4, we have also evaluated $T=0$ energies and entropies;
cf. Tab. 1 and 2. The results show that the 1RSB and 2RSB approximations are
still not thermodynamically acceptable at very low temperatures near the
respective $\alpha_c$, because the $T=0$ entropies turn out to be negative,
which is strictly forbidden in a system with discrete variables.

Internal energy $u$ and entropy $s$ are computed from the relations
$u = {\partial \beta f\over \partial \beta}$ and $s=\beta\,u-\beta\,f$ in units
of $k_{\hbox{\mf B}}$. The computations are straightforward, if perhaps tedious
in details. For the $T=0$ entropy, one obtains
 \begin{eqnarray}
s\,\,_{T=0}=-{\alpha\over  2 }\left( \,{C\over 1-C}+\ln  (1-C) \right) ,
\end{eqnarray}
which is {\em formally independent} of the degree of approximation in the
finite-step RSB scheme, provided the response parameter $C$ is defined as
$C=\beta\,(\hat q-q(1))$, with $\hat q$ the diagonal entry in the matrix of
Edwards-Anderson order parameters, and $q(1)$ the off-diagonal entry in the
{\em innermost blocks}, that is $q(1)=q$ in the RS approximation, and
$q(1)=q_k$ in
the kRSB scheme, $k=1,2,\ldots$. Thus as long as $C(T=0)$ is non-zero, the
$T=0$ entropy will come out negative.

The same {\em formal independence} for the zero entropy expression of the
degree of approximation in the Parisi scheme is observed for the SK-model,
where
\begin{eqnarray}
s\,\,_{T=0}^{\hbox{\mf SK}}=-{C^2\over  4 }  ,
\end{eqnarray}
with $C=\beta\,(\hat q-q(1))=\beta\,(1-q(1))$, and $q(1)=q_k$ in the kRSB
scheme.

As a consequence, it is to be expected that no finite approximation in Parisi's
approximating scheme will yield thermodynamically acceptable solutions at $T=0$
for the retrieval phases of generalized Hopfield models, just as in the case of
the SK-model, where only the full hierarchical scheme of infintely many levels
of RSB gives a Parisi function $q(x)$ that is sufficiently smooth on the
$D=\beta\, x$ scale to produce a vanishing $C$ in the $T=0$ limit, and thereby
a vanishing zero temperature entropy.

Further evidence for the analogy between the retrieval phases of generalized
Hopfield models and the magnetic phase of the SK-model with ferromagnetic
anisotropy $\langle J_{ij}\rangle = J_o /N$ comes from comparing lines of
constant magnetization in a RS approximation; see Fig. 8.

\centerline{ }
\centerline{Fig. 8}
\centerline{ }

As in the Hopfield model, reentrance is observed in the RS approximation of the
SK-model phase diagram. Moreover, in {\em both} models, the AT-line is seen to
intersect the constant--magnetization lines slightly above the temperature
where they begin to bend back (to larger $J_0$ in the SK-model, to smaller
$\alpha$ in
the Hopfield model). Thus, reentrant behaviour as observed in RS approximations
is found to be AT unstable in both models, in a strikingly similar
fashion.\footnote{There is a mapping of the replica symmetric Hopfield model
onto the replica symmetric SK-model; for details see [16].}

Now according to the Parisi-Toulouse hypothesis [13], one effect of RSB in the
SK-model is, roughly, to {\em freeze} the value of the magnetization as a
function of temperature. That is, in the full RSB solution of the SK-model, the
iso--magnetization lines in the phasediagram 7 will be {\em verticals} below
the AT line. This statement is believed to be exact for the $m=0$ line, i.e.,
the phase boundary, and to constitute a very precise approximation otherwise
[13].

By analogy, and in view of the great similarity of the analytic structure, the
same is expected to hold in the case of generalized Hopfield models. As a
consequence, the retrieval phase boundary --- as the envelope of iso--overlap
lines with non-zero $m^\nu$ --- should in the full RSB scheme turn out to be
vertical (or very close to a vertical) below the point where the AT line
touches the RS phase boundary.

This hypothesis is {\em completely in accord} with the results of our 1RSB and
2RSB analy\-ses, which showed that the $T=0$ transition point $\alpha_c$ is
shifted to slightly higher values, closer to the point where they are expected
to be if the hypothesis were true, and never beyond the abscissa
$\alpha^{\hbox{\mf RS}}(T)_{\hbox{\mf max}}$ of the reentrant point, in
contrast to previously reported results [3].

\section{Summary and Discussion}

We have studied effects of RSB in generalized Hopfield--type models of
attractor neural networks. We have obtained 1RSB and 2RSB corrections to
RS results, which are {\it much}\/ smaller than expected from simulation
results [7--9] or previously reported for the standard model at $\bar a = 1/2$
[3]. In all cases the 1RSB and 2RSB storage capacities obtained were found
to be smaller than those attainable in the RS approximation at finite
temperature. Our results were found to be consistent with what is to be
expected if the Parisi--Toulouse hypothesis about the nature of the Parisi
function $q(x)$ [11,13] would hold for Hopfield--type models in the same manner
as it does in the SK spin glass model. On the basis of this hypothesis, we
conjecture that the reentrance phenomenon observed in RS analyses of
Hopfield--type models would simply disappear in a full hierarchical Parisi
RSB solution of these models.

This state of affairs raises the question of how to reconcile discrepancies
between theory and numerical experiments in these systems. Two possible
explanations come to mind, and both are, we think, worth checking.

One possibility is that $T=0$--Monte Carlo Dynamics gets trapped in energy
valleys which are surrounded by nonextensive energy barriers, $\Delta E
\sim N^\nu$ with $0 \le \nu < 1$. The existence of such nonextensive energy
barriers between thermodynamically unstable retrieval states and the spin glass
phase might well invalidate conventional (exponential) finite--size--scaling
expressions for first order phase transitions, on which the analyses
of simulation data [7,9] were based. Here we should, however, remark that
M\"uller [17] --- knowing of our results --- has recently performed Monte Carlo
simulations which would confirm our values for $\alpha_c$ on the basis of a
conventional finite--size--scaling analysis of his data.

A second possibility concerns the existence of dynamically frozen phases not
detectable in equilibrium treatments. Discrepancies analogous to those between
our results and those of simulations have, indeed, recently been observed
in the case of the binary perceptron [18], the $p$--spin interaction spin glass
[19], and in the case of fluctuating manifolds in random media [20], where
dynamically frozen phases
were observed in regions of parameter space in which static approaches yielded
ergodic phases.\\

{\bf Acknowledgements}
This work has been part of the PhD thesis of HS. Numerous fruitful
discussions with H. Horner are gratefully and with pleasure acknowledged.
RK thanks the Laboratoire de Physique Th\'eorique of ENS for the hospitality
extended to him while parts of this paper were being written.

\vfill\eject

\appendix
\section{The 1RSB Approximation}

In this appendix, we present the main ideas that go into the  evaluation
of the replica free energy (\ref{nf}) in the 1RSB approximation, and into the
derivation of the corresponding 1RSB version of the fixed point equations
(\ref{Sattel}). The 1RSB approximation is based on the ansatz (\ref{q1rsb}) for
the matrix of Edwards--Anderson order parameters, in which $m$ is a
partitioning
parameter for the set of $n$ replica that is to be determined from a
stationarity condition on $f$. Clearly, the structure of $\q$ is  inherited by
$\1-\beta\q$,
\begin{eqnarray}
\1-\beta\q=(1-\beta(\hat q-q_1))\,{\bf 1}_n -\beta(q_1-q_0)\,{\bf 1}_{n\over m}
\otimes{\bf e}_m{\bf e}_m^{\rm T} - \beta q_0\,{\bf e}_n{\bf e}_n^{\rm T},
\label{1mbq}
\end{eqnarray}
as well as by $\r$ in virtue of (\ref{rSattel}). Using this structure, we have
to evaluate the various terms appearing in the free energy (\ref{nf}).

(i) In order to evaluate $\ln\det(\1-\beta\q)$, we diagonalize (\ref{1mbq}).
This is accomplished by noting that $\1_n = \1_{n\over m} \otimes \1_m$, and
${\bf e}_n{\bf e}_n^{\rm T}={\bf e}_{n\over m}{\bf e}_{n\over m}^{\rm T}\otimes
{\bf e}_m{\bf e}_m^{\rm T}$, and that matrices with tensor product structure
can be diagonalized separately in each tensor product component. Matrices of
the form ${\bf e}_k{\bf e}_k^{\rm T}$ have one eigenvalue $k$ and a
$(k-1)$--fold eigenvalue zero. Since they trivially commute with the
corresponding unit matrices $\1_k$, the full spectrum of (\ref{1mbq}) is
readily obtained to yield
\begin{eqnarray}
\det(\1-\beta\q)& = &\left(1-\beta(\hat q- q_1) -\beta m (q_1-q_0) -n
\beta q_0 \right) \nonumber\\
& & \left(1-\beta(\hat q- q_1) -\beta m (q_1-q_0) \right)^{{n\over
m}-1}\nonumber\\
& & \left(1-\beta(\hat q- q_1) \right)^{n-{n\over m}}.
\end{eqnarray}

(ii) Next, the term $\sum_{a\leq b} r_{ab}\,q_{ab}$ appearing in (\ref{nf})
is evaluated by endowing $\r$ with the same block structure as $\q$, and by
parametrizing it analogously. This gives
\begin{equation}
\sum_{a\leq b} r_{ab}\,q_{ab}= {1\over 2} \left(n \hat r \hat q + \sum_{a,b}
r_{ab}\,q_{ab}\right) = {n\over 2} \big [\, 2 \hat r \hat q + (m-1) r_1 q_1 +
(n-m) r_0 q_0 \, \big]
\end{equation}

(iii) In view of (\ref{rSattel}), to express the elements of $\r$ in terms of
those of $\q$, we have to invert $\1-\beta\q$. This is done by noting that
the matrices $\1_k$ and ${\bf e}_k{\bf e}_k^{\rm T}$ form a closed algebra,
since ${\bf e}_k{\bf e}_k^{\rm T}\, {\bf e}_k{\bf e}_k^{\rm T} = k
{\bf e}_k{\bf e}_k^{\rm T}$. Thus the inverse of $\1-\beta\q$ must be of the
same structure as $\1-\beta\q$ itself, albeit with different coefficients in
front of the three tensor product matrices appearing in (\ref{1mbq}). These
coefficients are computed from the condition that the product of $\1-\beta\q$
and its inverse should give a unit matrix. This yields (\ref{r1rsb}), with the
$Q_{q_i}$ defined by (\ref{Qi1rsb}).

(iv) Finally, to evaluate the single--site replica partition function
corresponding to the Hamiltonian (\ref{hrep}), we have to decouple the
replicated spins coupled through the term $\sum_{a\leq b} r_{ab}  s^{a} s^{b}$.
With $\r$ of 1RSB form analogous to (\ref{q1rsb}), we get
\begin{equation}
\sum_{a\leq b} r_{ab} s^{a} s^{b} = {1\over 2} \left[(2\hat r -r_1)
\sum_{a=1}^n
(s^a)^2 + (r_1-r_0)\sum_{k=1}^{n/m} \left(\sum_{a=1}^m s^{(k-1)m+a}\right)^2
+r_0 \left( \sum_{a=1}^n s^a \right)^2 \right]\ .
\label{1rabsab}
\end{equation}
Gaussian linearization of the last term in (\ref{1rabsab}), using a Gaussian
variable $z$, will produce $n/m$ identical uncoupled blocks of size $m$ in the
evaluation of (\ref{zrep}). Within these blocks there is still a coupling
between spins due to the next--to--last term in (\ref{1rabsab}). These are
decoupled using a Gaussian $z_1$. This then yields
\begin{equation}
\langle \ln\hat Z \rangle_{\xi^\nu}= n \left \langle \ln \langle \tilde Z^m
\rangle_{z_1}^{{1\over m}} \right\rangle_{z,\xi^\nu}
\end{equation}
in (\ref{nf}), as $n\to 0$, with $\tilde Z$ the partition function
corresponding
to the on--site Hamiltonian (\ref{1rsbH}). Similar linearization techniques
are readily seen to produce the 1RSB fixed point equations (\ref{1rsbfpe}) from
(\ref{Sattel}), as the $n\to 0$--limit is taken.

Collecting all items so far computed, we obtain the 1RSB expression
(\ref{f1rsb32}). Stationarity of this expression with respect to the
partitioning parameter $m$ requires (\ref{1RSBHOPF}) to hold, which completes
our derivation of the 1RSB approximation.

\section{The 2RSB Approximation}

In the present appendix, we sketch the evaluation of (\ref{nf}) in the 2RSB
approximation, and the derivation of the corresponding 2RSB version of the
fixed point equations (\ref{Sattel}). The following outline completely
parallels
that of Appendix A, and we will not repeat the arguments in detail.

The 2RSB approximation is based on the ansatz (\ref{q2rsbann}) for
the matrix of Edwards--Anderson order parameters, in which $m_1$ and $m_2$ are
partitioning parameters which are to be determined from a stationarity
condition on $f$. Again, the structure of $\q$ is  inherited by $\1-\beta\q$,
\begin{eqnarray}
\1-\beta\q & = & (1-\beta(\hat q-q_2))\,{\bf 1}_n
-\beta (q_2-q_1)\,{\bf 1}_{n\over m_2}\otimes{\bf e}_{m_2}{\bf e}_{m_2}^{\rm T}
\nonumber\\
& & - \beta (q_1-q_0)\,{\bf 1}_{n\over m_1} \otimes {\bf e}_{m_1}{\bf
e}_{m_1}^{\rm T} - \beta q_0\,{\bf e}_n{\bf e}_n^{\rm T}.
\label{2mbq}
\end{eqnarray}
as well as by $\r$ in virtue of (\ref{rSattel}).

(i) In order to evaluate $\ln\det(\1-\beta\q)$, we diagonalize (\ref{2mbq})
along the lines outlined in Appendix A. This yields
\begin{eqnarray}
\det(\1-\beta\q)& = &\left(1-\beta(\hat q- q_2) -\beta m_2 (q_2-q_1)
- \beta m_1 (q_1 - q_0) - \beta n q_0 \right) \nonumber\\
& & \left(1-\beta(\hat q- q_2) -\beta m_2 (q_2-q_1)
- \beta m_1 (q_1 - q_0) \right))^{{n\over m_1}-1} \nonumber\\
& & \left(1-\beta(\hat q- q_2) -\beta m_2 (q_2-q_1) \right))^{{n\over m_2} -
{n\over m_1}} \nonumber\\
& & \left(1-\beta(\hat q- q_1) \right)^{n-{n\over m_2}}.
\end{eqnarray}

(ii) The term $\sum_{a\leq b} r_{ab}\,q_{ab}$ appearing in (\ref{nf})
is evaluated by endowing $\r$ with the same block structure as $\q$, and by
parametrizing it analogously. This gives
\begin{eqnarray}
\sum_{a\leq b} r_{ab}\,q_{ab}  = {n\over 2} \big[\, 2 \hat r \hat q + (m_2-1)
r_2 q_2 + (m_1 - m_2) r_1 q_1 + (n - m_1) r_0 q_0 \, \big]
\end{eqnarray}

(iii) To express the elements of $\r$ in terms of those of $\q$, we have to
invert $\1-\beta\q$. This is done as outlined in Appendix A, and yields
(\ref{r2rsb}), with the $Q_{q_i}$ defined by (\ref{Qi2rsb}).

(iv) Next, to evaluate the single--site replica partition function
corresponding
to the Hamiltonian (\ref{hrep}), we have to decouple the replicated spins
coupled through the term $\sum_{a\leq b} r_{ab}  s^{a} s^{b}$. With $\r$ of
2RSB form analogous to (\ref{q2rsbann}), we get
\begin{eqnarray}
\sum_{a\leq b} r_{ab} s^{a} s^{b} & = & {1\over 2} \Bigg [ (2\hat r -r_2)
\sum_{a=1}^n (s^a)^2 + (r_2-r_1)\sum_{k=1}^{n/m_2} \Big(\sum_{a=1}^{m_2}
s^{(k-1)m_2 +a} \Big)^2 \nonumber\\
& & \ \ + (r_1-r_0)\sum_{k=1}^{n/m_1} \Big(\sum_{a=1}^{m_1}
s^{(k-1)m_1 +a} \Big)^2 + r_0 \Big( \sum_{a=1}^n s^a \Big)^2 \Bigg ]\ .
\label{2rabsab}
\end{eqnarray}
This structure suggests an iterative Gaussian linearization scheme as in the
case of the 1RSB approximation dicussed in the previous appendix. A Gaussian
$z$ is introduced to decouple spins in different blocks of size $m_1$, and
creates $n/m_1$ identical independent, i.e., uncoupled blocks of this size.
Within a block of size $m_1$, there are $m_1/m_2$ identical blocks of size
$m_2$, which are  decoupled through a Gaussian $z_1$. Finally the $m_2$ spins
within each of these smaller blocks are decoupled using a Gaussian $z_2$. This
results in
\begin{equation}
\langle \ln\hat Z \rangle_{\xi^\nu}= n \left \langle \ln \left\langle \langle
\tilde Z^{m_2} \rangle_{z_2}^{{m_1\over m_2}} \right\rangle_{ z_1}^{{1\over
m_1}} \right\rangle_{z,\xi^\nu}
\end{equation}
in (\ref{nf}), as $n\to 0$. Here $\tilde Z$ is the partition function
corresponding to the on--site Hamiltonian (\ref{2rsbH}). Again, similar
linearization techniques are seen to produce the 2RSB fixed point equations
(\ref{2rsbfpe}) from (\ref{Sattel}), as the $n\to 0$--limit is taken.

Collecting all items, we obtain the 2RSB expression (\ref{f2rsb}). Stationarity
of this expression with respect to the partitioning parameters $m_1$ and $m_2$
requires (\ref{stat1}) and (\ref{stat2}) to hold, which completes
our derivation of the 2RSB approximation.


\begin{thebibliography}{888888888}

\bibitem [1] {Ho89}
H. Horner, Z. Phys. B {\bf 75}, 133 (1989)

\bibitem [2] {PVAm89}
C.J. Perez--Vicente and D.J. Amit, J. Phys. A {\bf 22}, 559
(1989)

\bibitem [3] {CAG86}
A. Crisanti, D.J. Amit, and H. Gutfreund, Europhys. Lett. {\bf 2}, 337
(1986)

\bibitem [4] {NaCa91}
J.P. Naef and A. Canning, J. Phys. {\bf 2}, 247 (1992)

\bibitem [5] {Hop82}
J.J. Hopfield, Proc. Natl. Acad. Sci. USA {\bf 79}, 2554 (1982)

\bibitem [6] {KiSh78}
D. Sherrington  and  S. Kirkpatrick, Phys. Rev. Lett. {\bf 35}, 1792 (1975);
S. Kirkpatrick and  D. Sherrington, Phys.  Rev. {\bf B 17}, 4384 (1978); see
also M. M\'ezard, G. Parisi, and M. A. Virasoro, {\em Spin Glass Theory and
Beyond}, (World Scientific, Singapore, 1987)

\bibitem [7] {AGS87} D.J. Amit, H. Gutfreund, and H. Sompolinsky,
Ann. Phys. (N.Y.) {\bf 173}, 30 (1987)

\bibitem [8] {Koh90}
G.A. Kohring, J. Stat. Phys. {\bf 59}, 1077 (1990)

\bibitem [9] {FoLo92}
B.M. Forrest und A. Loettgers, Z. Phys. B {\bf 88}, 309
(1992)

\bibitem [10] {Pa80a}
 G. Parisi, J. Phys. {\bf  A13}, L115
(1980)

\bibitem [11] {Pa80b}
 G. Parisi, J. Phys. {\bf  A13}, 1101
(1980)

\bibitem [12] {Huy93}
J. Huyghebaert, private Communication; see also D. Boll\'e and J. Huyghebaert,
{\em Mixture states and the storage of biased patterns in the Hopfield model:
a replica--symmetry breaking solution}, preprint (Leuven, 1993) KUL--TF--93/46

\bibitem [13] {PaTo80}
G. Parisi  and  G. Toulouse, J. Phys. (Paris) Lett. {\bf 41},  L361 (1980);
J. Vannimenus, G. Toulouse, and G. Parisi, J. Phys. (Paris)   {\bf 42}, 565
(1981)

\bibitem [14] {Gar87}
E. Gardner, Europhys. Lett. {\bf 4}, 481 (1987); ---, J. Phys. A {\bf 21},
257--270 (1988)

\bibitem [15] {dATh78}
 J.R.L. de Almeida and  D.J. Thouless, J. Phys. {\bf A11}, 983 (1978)

\bibitem [16] {St93}
H. Steffan, {\em Replikasymmetriebrechung in Attraktor-Neuronalen-Netzen},
PhD--Thesis, Heidelberg (1993), unpublished

\bibitem [17] {Muel93}
K.R. M\"uller, private communication (1993)

\bibitem [18] {Ho92}
H. Horner, Z. Phys. B {\bf 86} 291 (1992)

\bibitem [19] {CHS93}
A. Crisanti and H.J. Sommers,  Z. Phys. B {\bf 87} 341 (1992);
A. Crisanti, H. Horner, and H.J. Sommers, Z. Phys. B {\bf 92} 257 (1993)

\bibitem [20] {KiHo93}
M. M\'ezard and G. Parisi, J. Phys.  {\bf A23}, L1229 (1990); ---, J. Phys. I
France {\bf 1} 809 (1991); H. Kinzelbach and H. Horner, J. Phys. I France {\bf
3} 1329 (1993); --- ibid. {\bf 3} 1901 (1993)

\end{thebibliography}
\end{document}